# A multiscale-multiphysics strategy for numerical modeling of thin piezoelectric sheets

Claudio Maruccio[1*], Laura De Lorenzis[2], Luana Persano[3] and Dario Pisignano[3,4]

[1]*Department of Innovation Engineering, University of Salento, via Monteroni, Lecce, Italy,*
*claudio.maruccio@unisalento.it*

[2]*Institut für Angewandte Mechanik, Technische Universität Braunschweig, Germany,*
*l.delorenzis@tu-braunschweig.de*

[3]*National Nanotechnology Laboratory, CNR-Istituto Nanoscienze, Lecce, via Arnesano, Italy,*
*luana.persano@nano.cnr.it*

[4] *Department of Mathematics and Physics "E. De Giorgi", University of Salento, via Monteroni, Lecce, Italy,*
*dario.pisignano@unisalento.it*

*Abstract* - **Flexible piezoelectric devices made of polymeric materials are widely used for micro- and nano-electro-mechanical systems. In particular, numerous recent applications concern energy harvesting. Due to the importance of computational modeling to understand the influence that microscale geometry and constitutive variables exert on the macroscopic behavior, a numerical approach is developed here for multiscale and multiphysics modeling of piezoelectric materials made of aligned arrays of polymeric nanofibers. At the microscale, the representative volume element consists in piezoelectric polymeric nanofibers, assumed to feature a linear piezoelastic constitutive behavior and subjected to electromechanical contact constraints using the penalty method. To avoid the drawbacks associated with the non-smooth discretization of the master surface, a contact smoothing approach based on Bézier patches is extended to the multiphysics framework providing an improved continuity of the parameterization. The contact element contributions to the virtual work equations are included through suitable electric, mechanical and coupling potentials. From the solution of the micro-scale boundary value problem, a suitable scale transition procedure leads to the formulation of a macroscopic thin piezoelectric shell element.**

I. INTRODUCTION

The discovery of appreciable piezoelectricity on ceramic and polymeric materials such as ZnO and Polyvinylidene fluoride (PVDF) and its copolymers has led to the development of a series of Nano Wire-based piezoelectric nanogenerators [1]. Most recent models have showed the capability of powering small electronic devices. In particular, PVDF is a polymeric piezoelectric material with good piezoelectric and mechanical properties. Its piezoelectric coefficient is around -30 pC/N, which is higher than that of ZnO. Its flexible polymeric nature allows very high strains to be applied to a PVDF beam/sheet, thus high piezoelectric potentials can be expected. Far-field electrospinning was recently used [2] to produce piezoelectric sheets of PVDF nanofibers directly patterned onto the substrate (Fig.1). The stretching force due to the strong electric field applied between the nozzle and the substrate surface can pole the PVDF nanofibers into the β phase, which has a resulting component of the global polarization along the longitudinal direction of the resulting fibers. In previous experiments, the PVDF sheet was directly applied between two electrodes on a flexible plastic substrate for piezoelectric output measurement [2] .

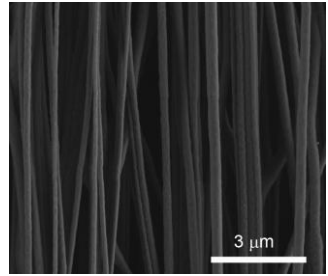

Fig.1. Scanning electron micrograph of a piezoelectric PVDF sheet.

The objective of this paper is to model the behavior of the aforementioned PVDF sheets made of polymeric nanofibers by a multiscale and multiphysics approach. This requires the definition of a representative volume element (RVE) at the microscale, the formulation and solution of a microscale boundary value problem (BVP), and the development of a suitable micro-macro scale transition. Due to the small thickness, $h_M \ll l_M$, a discretization with shell elements at the macroscale is numerically efficient (Fig.2).

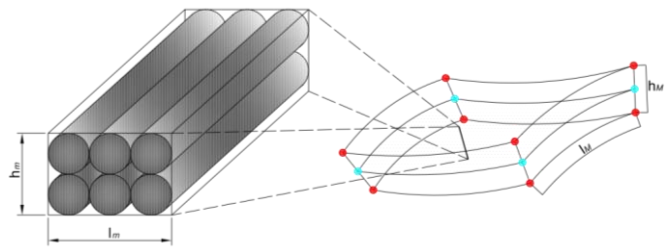

Fig.2. Determination of an RVE. $h_m$ and $l_m$ are the RVE thickness and length at the microscale, while $h_M$ and $l_M$ are the thickness and length at the macroscale, $l_m \ll l_M$.

The first part of the paper describes the kinematic behaviour of a piezoelectric shell following [3-5]. In the second part a microscale RVE element is defined and the theory of linear piezoelasticity [6] is briefly introduced along with its finite element formulation [7,8] at the microscale. Moreover some details about the formulation and implementation of suitable electromechanical contact elements using smoothing techniques [9,10] are provided. The third part describes numerical homogenization procedures for piezoelectric nanocomposites [11-15]. A multiscale approach is introduced to perform the scale transition between the micro- and macroscales. Details regarding the linearization of the finite element equations and implementation of this multiscale formulation are also provided. Finally, based on the presented multiscale and multiphysics framework the RVE geometry is analyzed to determine the effective material properties suitable for a macroscopic shell element definition. Advanced symbolic computational tools available in the AceGen-AceFem finite element environment within Mathematica [16,17] are used throughout this work, with the advantage that the tasks related to the finite element implementation are largely automated.

## II. MACROSCALE FORMULATION

Based on the analysis of the shell kinematics, the Green-Lagrangean strains and the electric field components in convective coordinates can be arranged in a generalized strain column vector:

$$\boldsymbol{\varepsilon} = [\varepsilon_{11}, \varepsilon_{22}, 2\varepsilon_{12}, \kappa_{11}, \kappa_{22}, 2\kappa_{12}, \gamma_1, \gamma_2, E_1, E_2, \varepsilon_{33}^0, \varepsilon_{33}^1, E_3^0, E_3^1]^T \quad \text{Eq.1}$$

where the membrane strain components $\varepsilon_{\alpha\beta}$ and the change of curvature $\kappa_{\alpha\beta}$ read:

$$\varepsilon_{\alpha\beta} = \frac{1}{2}(\boldsymbol{\psi}_{,\alpha} \cdot \boldsymbol{\psi}_{,\beta} - \boldsymbol{\psi}_{0,\alpha} \cdot \boldsymbol{\psi}_{0,\beta}); \quad \text{Eq.2}$$

$$\kappa_{\alpha\beta} = \frac{1}{2}(\boldsymbol{\psi}_{,\alpha} \cdot \boldsymbol{d}_{,\beta} + \boldsymbol{\psi}_{,\beta} \cdot \boldsymbol{d}_{,\alpha} - \frac{h_0}{2}\boldsymbol{\psi}_{0,\alpha} \cdot \boldsymbol{g}_{,\beta} - \frac{h_0}{2}\boldsymbol{\psi}_{0,\beta} \cdot \boldsymbol{g}_{,\alpha}); \quad \text{Eq.3}$$

In the previous equations the comma indicates partial derivation, Greek indices take the values 1, 2; $\xi_1, \xi_2$ are the natural coordinates of the shell middle surface, $\boldsymbol{\psi}, \boldsymbol{\psi_0}$ are respectively the current and initial position vectors of the shell middle surface, **g** and **a** are the initial and current normal (shell directors) and $h_0$ is the initial shell thickness. Moreover we introduced the quantity:

$$\boldsymbol{d} = \frac{h_0}{2}\lambda(\xi_1, \xi_2)\boldsymbol{a}(\xi_1, \xi_2); \quad \text{Eq.4}$$

where $\lambda(\xi_1, \xi_2)$ is the thickness stretch. The shear strain components take the form:

$$\gamma_\alpha = (\boldsymbol{\psi}_{,\alpha} \cdot \boldsymbol{d} - \frac{h_0}{2}\boldsymbol{\psi}_{0,\alpha} \cdot \boldsymbol{g}) \quad \text{Eq.5}$$

The electric components read:

$$E_\alpha = \frac{\partial \phi}{\partial \xi_\alpha}, \quad \text{Eq.6}$$

where $\phi$ is the electric potential. Furthermore, $\varepsilon_{33}^0, \varepsilon_{33}^1$ are the constant and linear components of the thickness strain and $E_3^0, E_3^1$ represent the constant and linear parts of the electric field along the thickness direction.

## III. MICROSCALE FORMULATION

At the microscale, the RVE consists in piezoelectric polymer fibers that feature a linear piezoelastic constitutive behavior and are subjected to electromechanical contact constraints. Piezoelectric problems are those in which an electric potential gradient causes deformation, and viceversa. The governing equations are the Navier equations and the strain-displacement relations for the mechanical field and Gauss and Faraday laws for the electrostatic field [6,7]. Moreover, the constitutive equations are:

$$a)\ T_{ij} = C_{ijkl}S_{kl} - e_{kij}E_k;\ b)\ D_i = e_{ikl}S_{kl} + \tilde{\epsilon}_{ik}E_k \quad \text{Eq.7}$$

where $C_{ijkl}$, $e_{ikl}$, and $\epsilon_{ik}$ are respectively the elastic, piezoelectric, and permittivity constants, whereas $S_{ij}, T_{ij}$ are the strain and stress components and $D_i, E_i$ are the electric displacement and the electric field components, respectively. Clearly, the coupling between mechanical and electric fields is determined by the piezoelectric coefficients. The 3D electromechanical frictionless contact element is implemented with the following main characteristics:

- the contact formulation is based on the master-slave concept;
- Bézier patches are used for smoothing of the master surface;
- The impenetrability condition is extended to the electromechanical setting by imposing equality of the electric potential in case of closed contact. The electromechanical constraints are regularized with the penalty method.

For each slave node, the normal gap is computed as:

$$g_N = (x_s - x_m) \cdot n \qquad \text{Eq.8}$$

where $x_s$ is the position vector of the slave node, $x_m$ is the position vector of its normal (i.e. minimum distance) projection point onto the master surface, and $n$ is the outer normal to the master surface at the projection point. The sign of the measured gap is used to discriminate between active and inactive contact conditions, a negative value of the gap leading to active contact. The electric field requires the definition of the contact electric potential jump:

$$g_\phi = (\phi_s - \phi_m) \qquad \text{Eq.9}$$

where $\phi_s$ and $\phi_m$ are the electric potential values in the slave node and in its projection point on the master surface. A tensor product representation of one-dimensional Bézier polynomials is used to interpolate the master surfaces in the contact interface for a three-dimensional problem according to the relation:

$$x(\xi_1, \xi_2) = \sum_{k=0}^{m} \sum_{l=0}^{m} d_{kl} B_k^m(\xi_1) B_l^m(\xi_2) \qquad \text{Eq.10}$$

where $B_k^m(\xi_1)$ and $B_l^m(\xi_2)$ are the Bernstein polynomials and $d_{kl}$ the coordinates of the control points. With the same procedure, to interpolate the electric potential on the master surface the following relation is introduced:

$$\phi(\xi_1, \xi_2) = \sum_{k=0}^{m} \sum_{l=0}^{m} \phi_{kl} B_k^m(\xi_1) B_l^m(\xi_2) \qquad \text{Eq.11}$$

where $\phi_{kl}$ is the potential evaluated at 16 control points as a function of the potential at the auxiliary points $\hat{\phi}_{kj}$ and at the master nodes $\phi_{ij}^2$. According to standard finite element techniques, the global set of equations can be obtained by adding to the variation of the energy potential representing the continuum behavior the virtual work associated to the electromechanical contact contribution provided by the active contact elements. If $\hat{u}$ is the set of degrees of freedom (DOF) used to discretize the displacement field $u = u(\hat{u})$, and $\hat{\phi}$ is the set of DOF used to discretize the electric potential field $\varphi = \varphi(\hat{\phi})$, so that $\hat{u} \cup \hat{\phi}$ is the vector of all nodal DOF, $\Pi_{global}$ is the global energy of the discretized system, the residual vector and the stiffness matrix terms resulting from the finite element discretization are determined according to:

$$\boldsymbol{R}_{u_i} = \frac{\delta \Pi_{global}}{\delta \hat{u}_i} \; ; \; R_{\phi_i} = \frac{\delta \Pi_{global}}{\delta \hat{\phi}_i} \; ;$$

$$\boldsymbol{K}_{uu_{i,j}} = \frac{\delta R_{u_i}}{\delta \hat{u}_i} \; ; \; K_{\phi\phi_{i,j}} = \frac{\delta R_{\phi_i}}{\delta \hat{\phi}_j} \; ;$$

$$\boldsymbol{K}_{u\phi_{i,j}} = \frac{\delta R_{u_i}}{\delta \hat{\phi}_j} \; ; \; \boldsymbol{K}_{\phi u_{i,j}} = \frac{\delta R_{\phi_i}}{\delta \hat{u}_j} \; . \qquad \text{Eq.12}$$

## IV. HOMOGENIZATION PROCEDURE

The main idea of homogenization is to find a globally homogeneous medium equivalent to the original composite, where the equivalence is intended in an energetic sense as per Hill's balance condition. Coupling between the macroscopic and microscopic scales is here based on averaging theorems, see [18,19]. Formulated for the electromechanical problem at hand, Hill's criterion in differential form reads:

$$\bar{T}_{ij}\delta\bar{S}_{ij} + \bar{D}_i\delta\bar{E}_i = \frac{1}{V}\iiint T_{ij}\delta S_{ij}dV + \frac{1}{V}\iiint D_i\delta E_i \, dV \qquad \text{Eq.13}$$

and requires that the macroscopic volume average of the variation of work performed on the RVE is equal to the local variation of work on the macroscale. In the previous equation: $\bar{T}_{ij}$, $\bar{S}_{ij}$, $\bar{D}_i$ and $\bar{E}_i$ represent respectively the average values of stress, strain, electric displacement and electric field components. Hill's lemma leads to the following equations:

$$\bar{T}_{ij} = \frac{1}{V}\iiint T_{ij}dV; \quad \bar{S}_{ij} = \frac{1}{V}\iiint S_{ij}dV;$$
$$\bar{D}_i = \frac{1}{V}\iiint D_i dV; \quad \bar{E}_i = \frac{1}{V}\iiint E_i dV \qquad \text{Eq.14}$$

Classically three types of boundary conditions are used for an RVE: prescribed displacements, prescribed tractions and periodic boundary conditions. Among them, periodic boundary conditions lead to a more reasonable estimation of the effective properties and therefore are often preferred. In this work, periodic boundary conditions are expressed as linear constraints and are implemented in Acegen as multipoint constraints using the Lagrange multiplier method. For simplicity, meshing of the RVE is performed uniformly such that identical nodes are present on all faces of the RVE. The final constitutive equation in the homogenized setting reads:

$$\begin{pmatrix}\bar{T}\\\bar{D}\end{pmatrix} = \begin{pmatrix}\bar{C} & -\bar{e}\\\bar{e} & \bar{\varepsilon}\end{pmatrix}\begin{pmatrix}\bar{S}\\\bar{E}\end{pmatrix} \qquad \text{Eq.15}$$

or in compact form:

$$\mathbb{T} = \mathbb{D}_{solid}^{macro}\mathbb{S} \qquad \text{Eq.16}$$

where $\bar{C}_{ijkl}$, $\bar{e}_{ikl}$, and $\bar{\varepsilon}_{ik}$ are respectively the homogenized elastic, piezoelectric, and permittivity constants. With some algebra and after integration on the shell thickness the previous equation becomes:

$$\mathbb{L} = \mathbb{D}_{shell}^{macro}\,\varepsilon \qquad \text{Eq.17}$$

with:

$$\mathbb{L} = [n_{11}, n_{22}, n_{12}, m_{11}, m_{22}, m_{12}, q_1, q_2, -d_1, -d_2, n_{33}^0, n_{33}^1, -d_3^0, -d_3^1]^T \qquad \text{Eq.18}$$

where $n_{\alpha\beta}$ are the membrane forces, $m_{\alpha\beta}$ are the bending moments, $q_\alpha$ are the shear forces and $n_{33}^0, n_{33}^1, -d_3^0, -d_3^1$ are the constant and linear components of membrane force and dielectric displacement in the thickness direction.

## V. RESULTS

If an RVE with PVDF fibers aligned in one direction is considered, the final material constitutive equations will be those of a transversely isotropic piezoelectric solid, and consequently the stiffness matrix, the piezoelectric matrix and the dielectric matrix will simplify so there remain in all 10 independent coefficients. In this work it is assumed that the elastic modulus $E = 150 N/mm^2$ and Poisson ratio $\nu = 0.29$, while the piezoelectric strain coefficients $d_{31}$, $d_{32}$, $d_{33}$ are equal to $20*10^{-12}$, $3*10^{-12}$, $-35*10^{-12}$ m/V, respectively, and the permittivity coefficients $\bar{\bar{\epsilon}}_{11}$, $\bar{\bar{\epsilon}}_{22}$, $\bar{\bar{\epsilon}}_{33}$ are all equal to $12\,\bar{\bar{\epsilon}}_0$, where $\bar{\bar{\epsilon}}_0 = 8.854*10^{-12}$ F/m is the void permittivity. According to the experimental results, each fiber in the RVE is assumed to have a radius $R = 1.0\mu m$. Hence, the RVE has a cubic geometry with a side length (L=2R) of 2 micron. For the evaluation of the effective properties, suitable boundary conditions have to be applied to the unit cell in such a way that, apart from one component of the strain/electric field vector, all other components are equal to zero [20, 21]. Then each effective coefficient can be easily determined by multiplying the corresponding row of the material matrix by the strain/electric field vector. In Fig.3 the mesh and the boundary conditions used during the analysis are illustrated. In particular different colors correspond to different degrees of freedom constrained during the analysis. Moreover, in figures 4 and 5 the contours of displacements and stresses are provided for boundary conditions allowing only stretching along the fiber longitudinal axis.

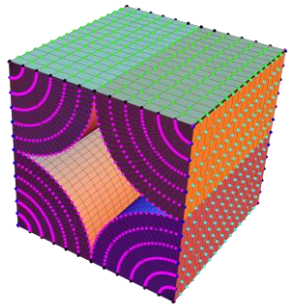

Fig.3. RVE mesh and boundary conditions.

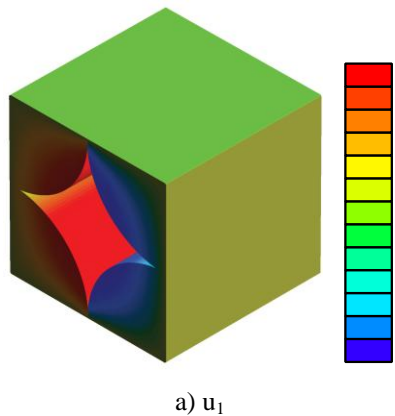

a) $u_1$

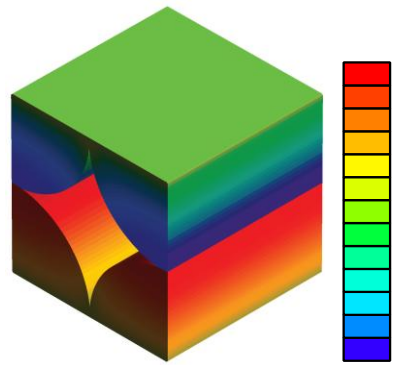

b) $u_2$

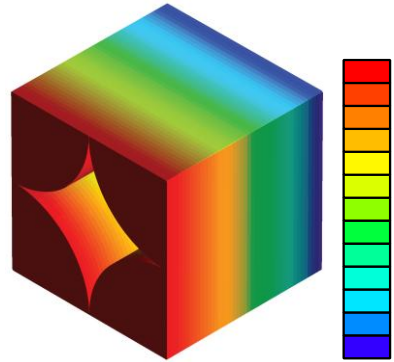

c) $u_3$

Fig.4. Microscale behavior, countour levels of displacement distribution in the RVE. Maximum and minimum corresponding to the vertical color scales are: a) +/- 0.104 [μm]; b) +/- 0.18 [μm]; c) -0.1/0 [μm].

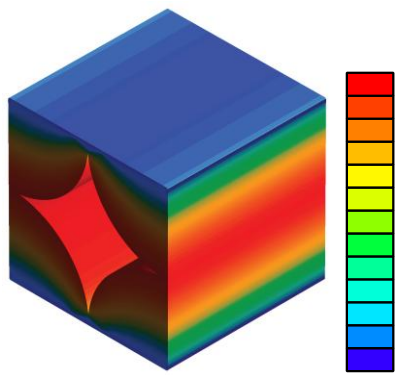

a) $T_{11}$

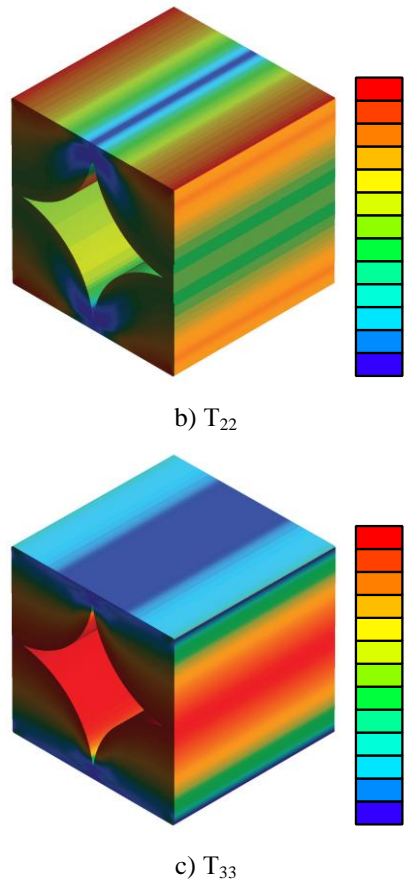

b) $T_{22}$

c) $T_{33}$

Fig.5. Microscale behavior, countour levels of stress distribution in the RVE. Maximum and minimum corresponding to the vertical color scales are: a) -2.39/+0.6 [N/mm$^2$]; b) -2.34/+0.48 [N/mm$^2$]; c) -6.43/-4.97 [N/mm$^2$].

Following the aforementioned procedure, it is possible to determine the evolution of the material coefficients under prescribed increasing displacements. The plots of the coefficients $D_{1,11}$, $D_{1,1}$ and $D_{2,2}$, $D_{1,2}$, $D_{11,11}$, are provided in Fig.6. Nonlinear behavior is due to the fiber interaction through electromechanical contact. In the figures, displacements are in micron and coefficients in MPa.

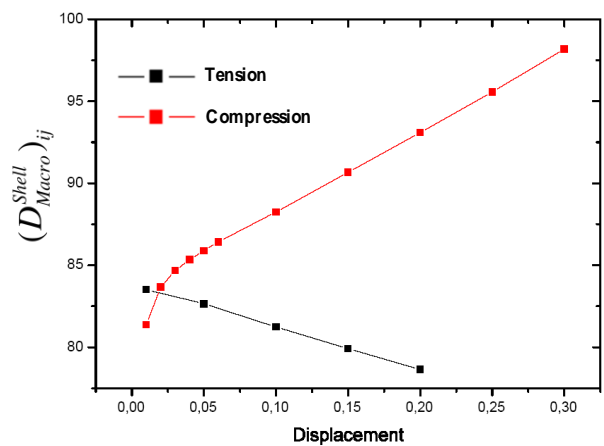

a) i=1=j=11

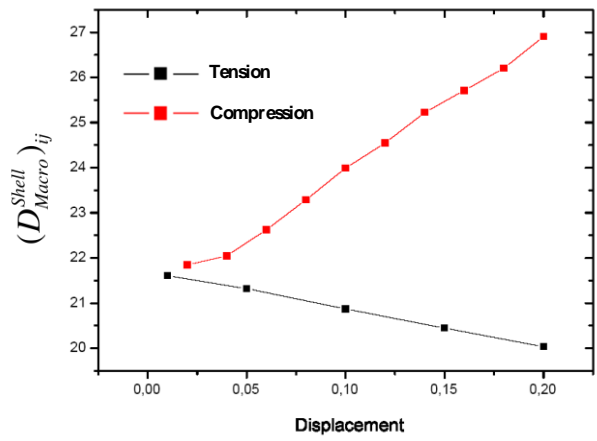

b) i=1 and j=2 or i=1 and j=2

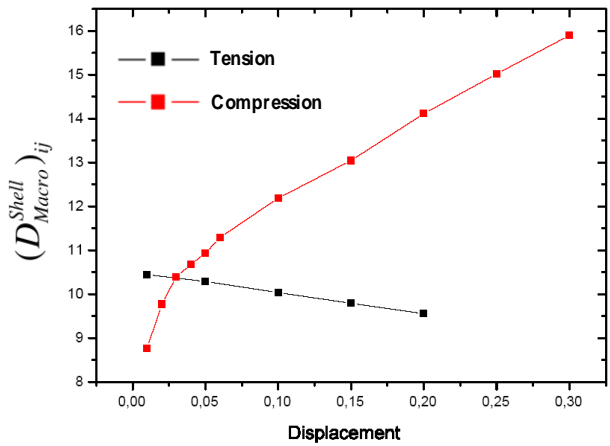

c) i=1 and j=11

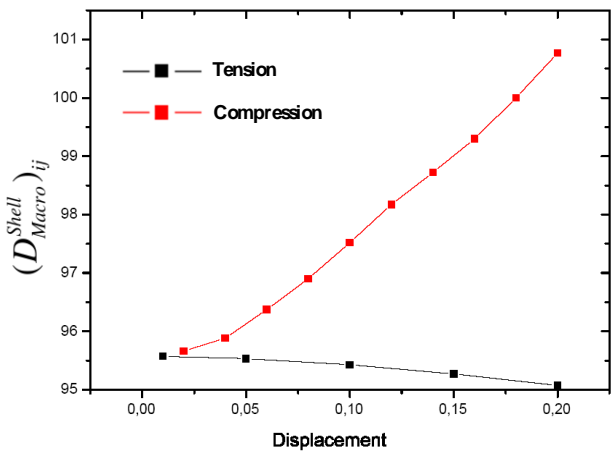

d) i=1 and j=1 or i=2 and j=2

Fig.6. Evolution of the effective coefficients of the shell: $D_{1,11}$, $D_{1,1}$ and $D_{2,2}$, $D_{1,2}$, $D_{11,11}$ as functions of the applied displacements. Coefficients are in N/mm$^2$ and displacements are in μm.

## VI. CONCLUSION

In this paper a multiscale and multiphysics homogenization method to model fibrous piezoelectric materials is presented. For each macroscale material point a fine scale boundary value problem is solved to determine the local material behavior on the macroscale. On the macroscopic level thin array of fibers are characterized by a large area to thickness ratio, such that a discretization with shell elements is numerically efficient. The material behavior is strongly influenced by the heterogeneous microstructure. Within homogenization based approaches the macroscopic constitutive behavior of the inhomogeneous material is modelled by means of an appropriate microscale RVE. The RVE consists in piezoelectric polymer fibers, assumed to feature a linear piezoelastic constitutive behavior and subjected to electromechanical contact constraints. A two-step homogenization scheme is applied to transfer the microscopic response to the macrolevel. Theoretical aspects are discussed and preliminary results for periodic structures are presented. Hence the computational homogenization procedure proposed is suitable for numerical modeling of flexible piezoelectric devices for energy harvesting technologies. Possible applications range from powering of remote sensors to extensive collection of energy. In the next future the development of these technologies will promote energy microgeneration and production of "Km 0 energy" in opposition to "macrogeneration" produced in massive plants that have a huge negative impact on the surrounding environment.

## ACKNOWLEDGEMENTS

Claudio Maruccio acknowledges the support from the Italian MIUR through the project FIRB Futuro in Ricerca 2010 Structural mechanics models for renewable energy applications (RBFR107AKG). Laura De Lorenzis, Dario Pisignano and Luana Persano acknowledge the support from the European Research Council under the European Union's Seventh Framework Programme (FP7/2007-2013), ERC Starting Grants INTERFACES (L. De Lorenzis, grant agreement n. 279439) and NANO-JETS (D. Pisignano and L. Persano, grant agreement n. 306357).